\documentclass[aip,rsi,preprint]{revtex4-1}
\usepackage{amsmath}
\usepackage{graphicx}
\usepackage{lmodern,textcomp}
\usepackage{threeparttable}
\usepackage[breaklinks,colorlinks,allcolors=blue]{hyperref}

\begin{document}
\title{Design and characterization of dielectric filled TM\texorpdfstring{$_{110}$}{110} microwave cavities for ultrafast electron microscopy}
\author{W. Verhoeven}
\author{J. F. M. van Rens}
\author{A. H. Kemper}
\author{E. H. Rietman}
\author{H. A. van Doorn}
\author{I. Koole}
\affiliation{Department of Applied Physics, Coherence and Quantum Technology Group, Eindhoven University of Technology, P.O. Box 513, 5600 MB Eindhoven, The Netherlands}
\author{E. R. Kieft}
\affiliation{Thermo Fisher Scientific, Achtseweg Noord 5, 5651 GG Eindhoven, The Netherlands}
\author{P. H. A. Mutsaers}
\author{O. J. Luiten}
\email{o.j.luiten@tue.nl}
\affiliation{Department of Applied Physics, Coherence and Quantum Technology Group, Eindhoven University of Technology, P.O. Box 513, 5600 MB Eindhoven, The Netherlands}
\date{\today}

\begin{abstract}
Microwave cavities oscillating in the TM$_{110}$ mode can be used as dynamic electron-optical elements inside an electron microscope. By filling the cavity with a dielectric material it becomes more compact and power efficient, facilitating the implementation in an electron microscope. However, the incorporation of the dielectric material makes the manufacturing process more difficult. Presented here are the steps taken to characterize the dielectric material, and to reproducibly fabricate dielectric filled cavities. Also presented are two versions with improved capabilities. The first, called a dual-mode cavity, is designed to support two modes simultaneously. The second has been optimized for low power consumption. With this optimized cavity a magnetic field strength of $2.84\pm0.07$~mT was generated at an input power of $14.2\pm0.2$~W. Due to the low input powers and small dimensions, these dielectric cavities are ideal as electron-optical elements for electron microscopy setups.
\end{abstract}

\maketitle

\section{Introduction}
\subsection{Ultrafast electron microscopy}

Since the formulation of Scherzer's theorem in 1936~\cite{Scherzer1936,Hawkes2009}, which states that rotationally symmetric electro- and magnetostatic lenses will always have positive spherical aberrations, researchers have proposed different methods of incorporating time-varying fields inside an electron microscope to correct for these aberrations. However, due to the difficulty in generating rapidly varying fields at that time, the actual current that could be transmitted was too low in these schemes to make them viable. In the 1970's, with the availability of microwave technology, interest in pulsed electron beams started to grow again. Oldfield et al.~\cite{Oldfield1976} and Ura et al.~\cite{Ura1973} both independently managed to implement a two step method to create and then bunch electron pulses at GHz repetition rates, achieving sub-picosecond pulse lengths~\cite{Ura1978}.

Nowadays, interest in microwave cavities is on the rise again with the introduction of time-resolved electron experiments. Inside an ultrafast electron microscope (UEM), sub-picosecond resolution can be added to powerful techniques such as direct electron imaging~\cite{Zewail2006,Flannigan2012}, diffraction~\cite{Siwick2003,Morimoto2018}, energy spectroscopy~\cite{Carbone2008,VanDerVeen2015}, and Lorentz microscopy~\cite{Feist2017,Berruto2018}. In these setups, microwave cavities can be used to improve the temporal resolution~\cite{Oudheusden2010}, as a streak camera~\cite{Musumeci2010,Maxson2017}, or for time-of-flight spectroscopy~\cite{Verhoeven2016,Verhoeven2018}.

The generation of electron pulses inside an UEM is typically done using photoemission. However, several blanking methods are emerging as an alternative, in which an electron beam is periodically deflected over a small slit or aperture~\cite{ThesisAdam,Qui2015,Weppelman2018}. These methods have the advantage that no alterations have to be made to the electron source. Recently, it has been shown both theoretically~\cite{ThesisAdam,VanRens2017} and experimentally~\cite{Verhoeven2018} that electron pulses can be created by deflecting a continuous beam over a slit with a TM$_{110}$ cavity while maintaining the emittance and energy spread of the continuous electron source. This is achieved by using a conjugate blanking scheme, resulting in high quality ultrashort pulses. Using state-of-the-art synchronization systems, the microwave signal can be synchronized to a laser pulse~\cite{Brussaard2013,Walbran2015,Siwick2017}, allowing for pump-probe experiments to be performed with a high resolution.

\begin{figure}
\includegraphics{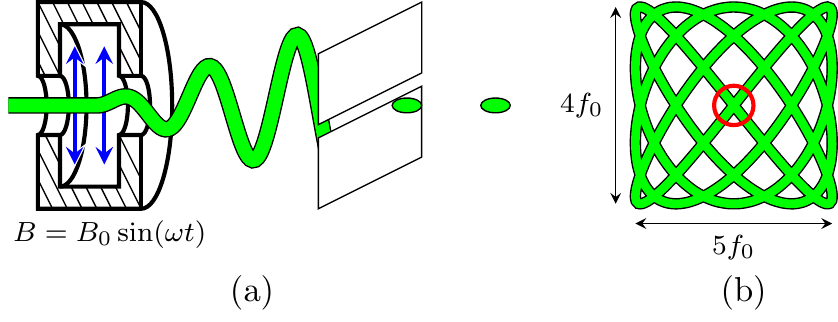}
\caption{(a) Principle of beam chopping using a cavity. The oscillating magnetic field periodically deflects the electron beam over a slit, resulting in ultrashort electron pulses. (b)~By deflecting the beam in perpendicular directions with two different frequencies both locked to a frequency $f_0$, the electron beam passes a pinhole (shown in red) at the difference frequency.}\label{fig:streaking}
\end{figure}

Microwave cavities are typically used at a frequency around 3 GHz due to the low cost of equipment operating in this frequency range. Since most pump-probe experiments use a lower repetition rate to allow for the sample to return into equilibrium, it has been proposed to deflect the beam in two directions at different harmonics of the desired repetition rate, allowing for electron pulses to be created at the difference frequency~\cite{ThesisAdam}. This principle is shown in Fig.~\ref{fig:streaking}(b), which schematically depicts an electron beam being deflected by two harmonics of a fundamental frequency $f_0$, resulting in a deflection following a Lissajous pattern. By placing an aperture in the center, shown here in red, pulses are created at the frequency $f_0$. By using a cavity that can support both frequencies, these Lissajous patterns can be created with a single cavity. This principle has also been experimentally demonstrated inside a UEM to reduce the repetition rate to 75~MHz using the 40$^\text{th}$ and 41$^\text{st}$ harmonic of a laser oscillator~\cite{VanRens2018}.

\subsection{Dielectric cavities}
Application of microwave cavities to manipulate electron beams at energies of 30--300~keV typically requires input powers in the range of $10^2$--$10^3$~W, for which expensive amplifiers are needed. Furthermore, this requires cooling systems with a high capacity to dissipate all the power, which can cause unwanted vibrations. Therefore, streak cavities can be filled with a dielectric material, reducing both the size and the power consumption of the cavity~\cite{Lassise2012}. This is shown in Table~\ref{table:cavities}, where the size and power consumption of vacuum and dielectric cavities are compared. Shown here are both the theoretical values of a cylindrical cavity, also called a pillbox cavity, and measurements performed on cavities that are optimized for power consumption. For the vacuum cavity, optimization is done by deviating from the pillbox geometry. This design is reported in Ref.~\citenum{ThesisThijs}. The dielectric cavity is optimized by optimizing the dielectric filling ratio, and is reported here.

\begin{table}
\begin{threeparttable}
\caption{Comparison of vacuum cavities and dielectric cavities at 3 GHz. Shown here are for both the theoretical values for a pillbox cavity, compared with measurements performed on optimized shapes.}\label{table:cavities}
\begin{tabular}{|l|cc|cc|}
\hline
&\multicolumn{2}{|l|}{vacuum cavity}&\multicolumn{2}{|l|}{dielectric cavity}\\
&pillbox$^\ast$&optimized&pillbox$^\ast$&optimized\\
\hline
radius (mm) & 60.98 & 66.20 & 10.16 & 36.80\\
$P/B_0^2$ ($10^6$ W\,T$^{-2}$) & 43.7 & $18.0\pm0.1$ & 3.85 & $1.67\pm0.04$\\
\hline
\end{tabular}
\begin{tablenotes}
\item $^\ast$Theoretical values from Eqs.~\eqref{eq:quality}--\eqref{eq:dielectriclosses} using $\sigma=5.8\times 10^7$ S/m, $\epsilon_r=36$ and $\tan \delta=1\times 10^{-4}$.
\end{tablenotes}
\end{threeparttable}
\end{table}

Although the design and implementation of a dielectric filled cavity was reported before in Ref.~\citenum{Lassise2012} as a proof-of-principle, this cavity proved to be difficult to reproduce. This makes it impossible to synchronize the microwave fields to a laser, since synchronization systems make use of components with a very narrow bandwidth~\cite{Brussaard2013,Walbran2015,Siwick2017}. Synchronization to a laser therefore requires well-defined resonant frequency of the cavity, deviating less than a few MHz from the intended value.

\begin{figure}
\includegraphics{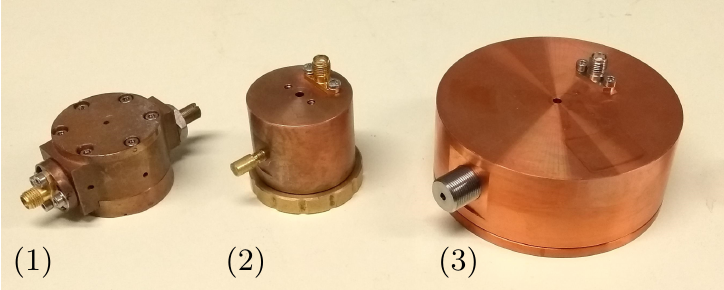}
\caption{From left to right: (1) the previous cavity design, (2) the current design using a screw cap, and (3) a low-loss cavity. All are filled with a dielectric material, and allow for operation at 3~GHz.}\label{fig:cavities}
\end{figure}

The difficulty in fabricating a dielectric cavity lies in the fact that the entire surface of the metal housing enclosing the dielectric material has to be electrically connected, but it also has to encompass the dielectric perfectly. If the housing is too loose, the dielectric can move, while a too tight housing will exert stress on the material, changing its properties. Both requirements cannot be met simultaneously for a housing with a fixed length. Therefore, the cavities reported here are closed using a screw cap, which ensures a good electric contact while the cap is tightened. Shown in Fig.~\ref{fig:cavities} are some cavity designs. From left to right are (1) the old design, (2) a screw cap housing which can either be used as a single or dual mode cavity, and (3) a design optimized for power consumption.

In this paper, the screw cap cavity design discussed in some detail. It is shown that control over the stress applied to the dielectric is crucial for a reproducible design. With this control, a cavity is fabricated for implementation in a 3~GHz UEM. Next, two more advanced cavities are presented. The first is a dual mode cavity, which can be used to create pulses at either 3~GHz or 75~MHz, and the second is a cavity optimized for low power consumption, allowing for larger field strengths to be generated.

\section{Theory}
\subsection{Uniformly filled cavity}
\begin{figure}
\includegraphics{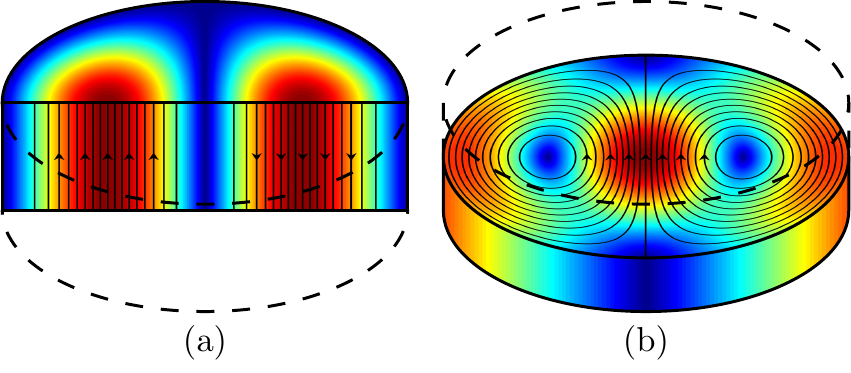}
\caption{(a) Longitudinal electric and (b) transverse magnetic field distribution of the TM$_{110}$ mode.}\label{fig:TM110}
\end{figure}
For a cylindrical cavity, the fields of an electromagnetic standing wave can be described in terms of the Bessel functions $J_n$ and $Y_n$. For a uniformly filled cavity, the field components of the TM$_{110}$ mode in cylindrical coordinates can be written as
\begin{align}
E_z&= \frac{2 \omega B_0}{k}J_1(kr)\cos(\theta)\cos(\omega t)\,,\nonumber\\
B_r&= \frac{2 B_0}{kr}J_1(kr)\sin(\theta)\sin(\omega t)\,,\\
B_\theta &= \frac{2B_0}{k}J_1'(kr)\cos(\theta)\sin(\omega t)\,,\nonumber
\end{align}
with $B_0$ the maximum magnetic field strength, $\omega$ the angular frequency, and $k=\sqrt{\mu\epsilon}\omega$ the wavenumber, with $\mu$ and $\epsilon$ the magnetic permeability and electric permittivity of the medium inside the cavity. The wavenumber and angular frequency of the mode are then given by the requirement that the electric field parallel to the metallic boundary is zero, \emph{i.e.}~$k=j_{1,1}/R$, with $R$ the radius of the cavity and $j_{m,n}$ the $n^\text{th}$ root of the $m^\text{th}$ Bessel function. Figure~\ref{fig:TM110} shows the field distribution of the TM$_{110}$ mode.

An important figure of merit for a resonant cavity is the quality factor $Q$, which relates the time-averaged energy $W$ stored in the cavity to the power loss $P_\text{loss}$. The quality factor is defined as
\begin{equation}
Q=\frac{\omega W}{P_\text{loss}}\,.\label{eq:quality}
\end{equation}
As the electric energy $W_e$ and magnetic energy $W_m$ stored at resonance are equal, the total energy stored in the volume $V$ of the TM$_{110}$ mode is given by~\cite{Pozar}
\begin{align}
W&=2W_e=\frac{1}{2}\int_V\epsilon_0\epsilon_r|\mathbf{E}|^2\,\text{d}^3x\label{eq:energy}\\
&=\frac{B_0^2\omega^2}{k^2}\epsilon_0\epsilon_r\pi R^2LJ_0^2(j_{1,1})\,.
\end{align}

Power losses arise from currents induced in the metallic walls by the magnetic field in the cavity. These losses are given by
\begin{align}
P_\text{surface}&=\frac{1}{2}\int_S\frac{|\mathbf{n}\times\mathbf{B}|^2}{\mu^2\sigma\delta_\text{skin}}\,\text{d}^2x\label{eq:metalliclosses}\\
&=\frac{2\pi B_0^2R}{\mu^2\sigma\delta_\text{skin}}(R+L)J_0^2(j_{1,1})\,,
\end{align}
where $\sigma$ is the conductivity of the wall, and $\delta_\text{skin}=\sqrt{\frac{2}{\mu\omega\sigma}}$ the skin depth.

By filling the cavity with a dielectric material, the total surface area decreases, resulting in a reduced power loss in the metallic walls. However, some additional power will also be dissipated by the dielectric material itself. Writing the permittivity as a complex number, power losses arise from a finite imaginary component~\cite{Griffiths}. Using the so-called loss tangent $\tan \delta=\text{Re}(\epsilon_r)/\text{Im}(\epsilon_r)$, the resulting dielectric loss is given by
\begin{align}
P_\text{volume}&=\frac{1}{2}\omega\epsilon_0\epsilon_r\tan\delta\int_V|\mathbf{E}|^2\,\text{d}^3x\label{eq:dielectriclosses}\\
&=\frac{\omega^3B_0^2}{k^2}\epsilon_0\epsilon_r\tan\delta\, \pi R^2LJ_0^2(j_{1,1})\,.
\end{align}
In order to reduce the total power consumption, the dielectric material therefore needs to have both a sufficiently large $\epsilon_r$ and a sufficiently low $\tan\delta$, such that the decrease in surface losses is greater than the increase in volume losses.

\subsection{Partially filled cavity}
\begin{figure}
\includegraphics{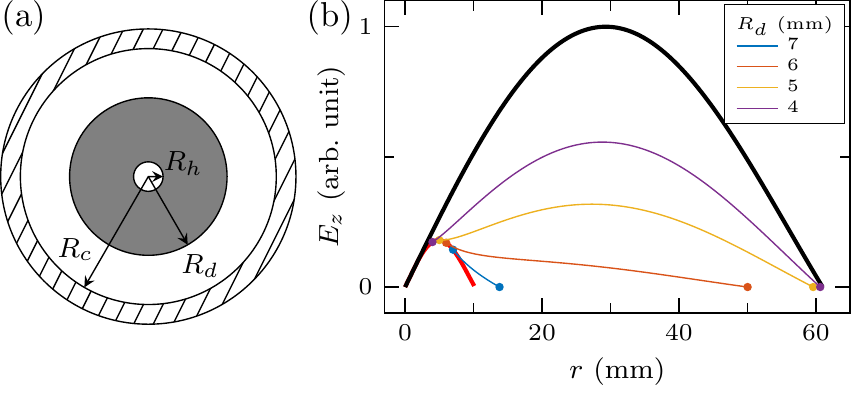}
\caption{(a)~Geometry of the partially filled cavity. A dielectric with radius $R_d$ is placed in a metallic cavity of radius $R_c$. In the center there is a hole with radius $R_h$. (b) Electric field as a function of radius for several partially filled cavities, with the circles at the start and end denoting $R_d$ and $R_c$. Also shown is the field of a vacuum cavity (black) and a dielectric cavity (red). The magnetic field strength at $r=0$ and the frequency are fixed for each curve.}\label{fig:partiallyFilledSolutions}
\end{figure}
In practice, the cavity cannot be filled uniformly, as electrons have to pass through the center. Furthermore, leaving space around the dielectric material allows for the insertion of an antenna and a tuning stub. In this section we will calculate the field distribution and power consumption using the geometry as shown in Fig.~\ref{fig:partiallyFilledSolutions}(b), where a dielectric of radius $R_d$ with a central hole of radius $R_h$ is placed in the center of a cavity with outer radius $R_c$.

For a non-uniform cylindrical cavity with a circularly symmetric cross-section, the field solutions can still be described in terms of the Bessel functions of the first and second kind $J_\nu$ and $Y_\nu$. For the TM$_{110}$ mode, the longitudinal electric fields in the three different regions are now given by
\begin{gather}
E_z(r,\theta)=2cB_0\cos(\theta)\times\nonumber \\ \left\{
\begin{array}{ll}
J_1(k_0r) & \text{if }r\leq R_h\,,\vspace{2pt}\\
\left[AJ_1(\sqrt{\epsilon_r}k_0r)+BY_1(\sqrt{\epsilon_r}k_0r)\right] & \text{if }R_h< r\leq R_d\,,\vspace{2pt}\\
\big[CJ_1(k_0r)+DY_1(k_0r)\big] & \text{if }R_d< r\leq R_c\,,\\
\end{array}\right.\,\label{eq:nonUniformSolution}
\end{gather}
where $k_0$ is the wave number in vacuum. This equation is subject to the boundary condition that the electric field parallel to the surface is continuous, i.e.\ that $E_z(R_h)$ and $E_z(R_d)$ are continuous. Furthermore, due to the absence of surface currents on the dielectric, the magnetic field is also continuous, implying that $E_z'(R_h)$ and $E_z'(R_d)$ are continuous. These four requirements determine the constants $A$, $B$, $C$, and $D$. Although analytical expressions exists for these constants~\cite{Carter2001}, they will not be given here. The wave number $k_0$ is again determined by the requirement that $E_z(R_c)=0$, and has to be found numerically.
Figure~\ref{fig:partiallyFilledSolutions}(b) shows the fields of Eq.~\eqref{eq:nonUniformSolution} for several values of $R_d$, together with the solutions of the vacuum cavity, and the dielectric cavity with a central hole $R_h=1.5$~mm. Here, both $B_0$ and $k_0$ are fixed, in which case also the constants $A$ and $B$ are fixed, resulting in the same field distribution within the dielectric. Colored circles in the graph show the positions of $R_d$ and $R_c$.

\begin{figure}
\includegraphics{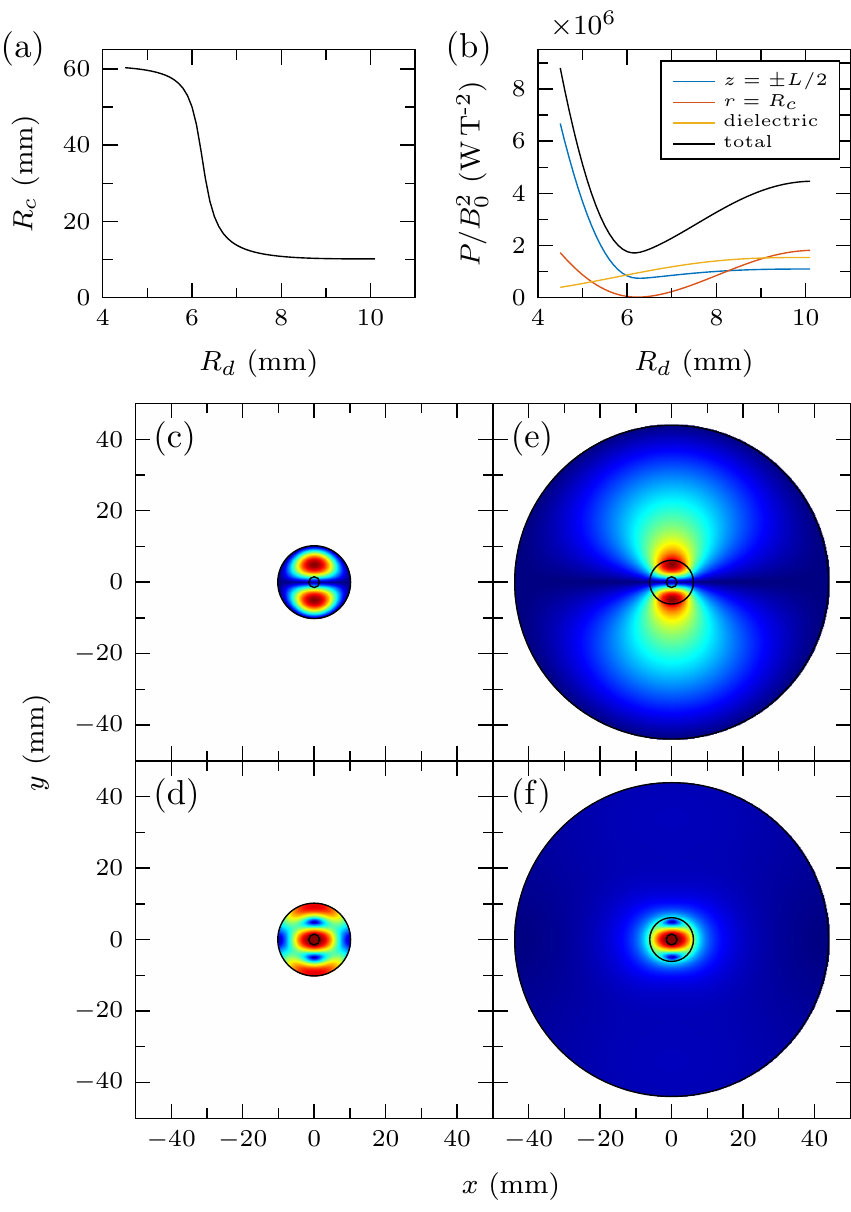}
\caption{(a) Cavity size and (b) power consumption per unit of magnetic field squared as a function of dielectric radius $R_d$ when keeping the resonant frequency fixed at 3 GHz. The power consumption is split into three contributions, namely surface losses at the top and bottom of the cavity ($z=\pm L/2$), surface losses at the side ($r=R_c$), and volume losses due to the dielectric. (c)~Electric and (d) magnetic field distribution of a dielectric cavity, compared to (e,f) those of an optimized cavity.}\label{fig:optimizationTheory}
\end{figure}

Varying $R_d$ and $R_c$ in such a way that the resonant frequency is kept constant allows for an interesting solution to be found, in which the electric field amplitude reaches its maximum inside the dielectric material, and slowly decays towards the wall, resulting in small magnetic fields outside the dielectric material. This can be seen from Fig.~\ref{fig:optimizationTheory} for the values of $R_d$ of 7 and 6. Because the magnetic field scales with the gradient of the electric field, these solutions will have smaller magnetic fields near the outside wall of the cavity, reducing the surface losses. An optimum therefore exists, where the outer radius of the cavity is small enough to suppress the standing wave outside the dielectric, but large enough to separate the walls from the large magnetic fields inside the dielectric.

The effect of varying $R_d$ on both $R_c$ and on the power consumption is shown in Figs.~\ref{fig:optimizationTheory}(a) and (b). The power consumption is split into three contributions, showing that the decrease is mostly due to the reduced losses at $r=R_c$. As the outer cavity radius $R_c$ approaches that of a vacuum cavity, losses go up again due to the emergence of strong magnetic fields outside the dielectric. Figures~\ref{fig:optimizationTheory}(c)--(f) show the electric and magnetic field distributions for both a completely filled dielectric cavity and a partial filled cavity with an optimized geometry. In all these figures, the dielectric material has been assumed to have $\epsilon_r=36$ and $\tan \delta=1\times 10^{-4}$, and the cavity length has been chosen to be $L=16.67$~mm. For these numbers, optimizing the filling ratio allows for a further reduction in power consumption by a factor 2.6. As a comparison, the optimized cavity requires an input power of 15.5~W to generate a field strength of 3~mT, whereas a vacuum pillbox equivalent would require 393~W.

\subsection{Electron beam dynamics}
When an electron beam moves through the center of the cavity, the electrons will experience a transverse Lorentz force. Assuming that the size of the electron beam is much smaller than that of the cavity, all electrons feel approximately the on-axis field, which is given by
\begin{equation}
\vec{B}=B_0 \sin(\omega t + \phi) \hat{y}\,,
\end{equation}
where a phase shift $\phi$ has been added between the arrival time of the electron and the fields in the cavity. This results in a Lorentz force acting on the electrons given by
\begin{equation}
\vec{F}_L=-q_ev_zB_0\sin(\omega t+\phi)\hat{x}\,,
\end{equation}
with $q_e$ the electron charge, and $v_z$ the longitudinal velocity. Assuming that the cavity has a top-hat field profile running from $-L/2<z<L/2$, and assuming that the change in longitudinal velocity inside the cavity is negligible, the total force exerted on the electrons causes a transverse change in velocity given by
\begin{equation}
v_x=-\frac{2q_ev_zB_0}{\gamma m_e\omega}\sin(\phi)\sin\left(\frac{\omega L}{2v_z}\right)\,,
\end{equation}
with $\gamma$ the Lorentz factor of the electrons, and $m_e$ the electron mass. A maximum deflection is achieved for a cavity length $L=\pi v_z/\omega$, \emph{i.e.}~when the electron transit time is half the oscillation period of the cavity. Placing a slit of width $s$ at a distance $d$, the resulting pulse length can then be approximated by
\begin{equation}
\tau=\frac{\gamma m_e s}{2|q_e|B_0d\sin(\omega L/2v_z)}\,.
\end{equation}
Given a slit of $s=10$ \textmu{}m placed at a distance of $d=10$~cm and electrons with $\gamma\approx1$, a field strength of $B_0=3$~mT is then required for pulses of $\tau\approx100$~fs.

\section{Cavity design}
\begin{figure}
\includegraphics{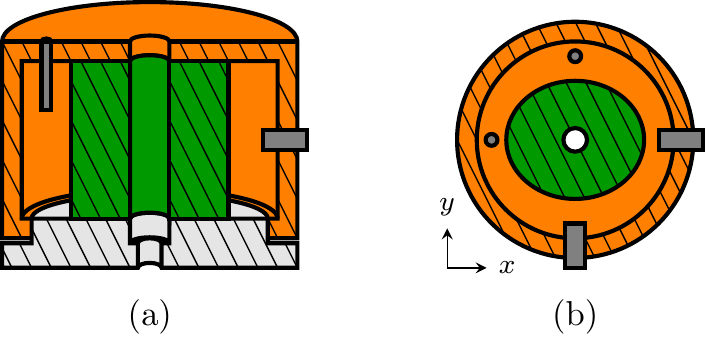}
\caption{(a) Cross section of the screw cap design, with the dielectric in dark green, copper housing in orange and screw cap in light gray, and a linear antenna and tuning stub in dark gray. (b)~Top view of the dual mode cavity, which is filled with an elliptical dielectric. Shown in dark gray are the antenna and tuning stub along both the short and long axes of the dielectric.}\label{fig:cavitySchematic}
\end{figure}

The cavities presented here are filled with a ZrTiO$_4$ ceramic doped with $<20\%$ SnTiO$_4$ produced by the company T-Ceram~\cite{TCeram}. The relative permittivity is quoted as $\epsilon_r=$36.5--38 in the frequency range of 0.9--18 GHz, with a loss tangent of $\tan \delta=2\times10^{-4}$ measured at 10 GHz.

Currently, the dielectric cavities are used both in a 200 keV UEM and a 30 keV electron beam setup. The cavity length is therefore chosen to be $L=16.67$~mm, close to the optimal length for 30 keV electrons. Although a longer cavity length would increase the deflection of 200 keV electrons, this is not necessarily beneficial for the quality of the electron pulses~\cite{VanRens2017}. Therefore, the reduction to $\sim68\%$ of the maximum deflection for 200 keV electrons is considered acceptable.

Figure \ref{fig:cavitySchematic}(a) shows a schematic cut-through of a cavity. Each cavity consist of a cylindrical dielectric placed inside a larger metal housing which is open from the bottom. Ample room is left around the dielectric to also insert an antenna and a tuning stub. As will be demonstrated in the next section, the ability to control the stress applied to the dielectric is of paramount importance. Therefore, the cavity is closed using a cap with screwing thread in the side which is tightened using a torque wrench, allowing for the stress to be applied controllably while the cap remains electrically connected. Cavities are cooled through water cooling that can be attached to either the housing or the screw cap. Power is coupled into the cavity using a linear antenna, and the frequency can be tuned several MHz using a metal tuning stub.

A hole in the center of 3~mm is used for the electrons to pass through. The housing, dielectric and cap are centered during assembly through these holes with a small rod. Furthermore, it is important to prevent electrons from hitting the dielectric to avoid charging. Therefore, the hole in the cap is narrowed to 2~mm on the outside. On the side of the dielectric the hole is kept 3~mm to avoid an asymmetry in the fringe fields at the entrance and exit.

Figure \ref{fig:cavitySchematic}(b) shows a top view of a dual mode cavity. Instead of a circular cylindrical dielectric, an elliptical cylinder is used to break the degeneracy of the differently oriented modes. Along the short and long axes an antenna and a tuning stub are placed, giving nearly independent control over each mode. In case of a single mode cavity, only one antenna and tuning stub is inserted.

For the cavities, a linear antenna is used. It can be shown that the field excited in the cavity by a linear antenna is given by~\cite{Collin}
\begin{equation}
e_0=-j\omega\mu_0\frac{\int_V\vec{J}(\vec{r})\cdot\vec{E}_n(\vec{r})\,\text{d}v}{k_0^2-k^2\left(1+\frac{1-j}{Q_0}\right)}\,,\label{eq:excitationAmplitude}
\end{equation}
where $e_0$ is the amplitude of the solenoidal electric mode $\mathbf{E}_n$ normalized by volume, which has a wavenumber $k_0$ and a quality factor $Q_0$, and $\mathbf{J}$ is the current distribution of the antenna. Furthermore, $\omega$ and $k$ are the angular frequency and wavenumber of the excitation signal. Although solving this equation relies on exact knowledge of the current distribution in the cavity and therefore the fields around the antenna, which are not known analytically, it shows that modes are mostly excited that have a high electric field at the antenna location.

Although the field strength per unit of current on the antenna is maximized by placing the antenna at the electric field maximum, this is not the ideal situation, as the cavity will likely become overcoupled, reducing the power transfer. The criterium for an antenna operating at high input power is therefore that its impedance is matched. Matching can be expressed in terms of the coupling factor $g$, which is given by
\begin{equation}
g=\frac{1-|\Gamma_v|}{1+|\Gamma_v|}\,,
\end{equation}
where $\Gamma_v$ is the voltage reflection coefficient. In practice, impedance matching is done by taking an antenna which is overcoupled, and then decreasing the length incrementally until the reflected power drops below a desired value. Once matched, the reflected power at resonance typically remains below a few percent during operation. Although in this design the antennas are placed outside the dielectric material where the electric field is low, overcoupling is relatively easy to achieve with a 50~$\Omega$ RF setup.

\section{Characterization}
\subsection{Screw cap design}
\begin{figure}
\includegraphics{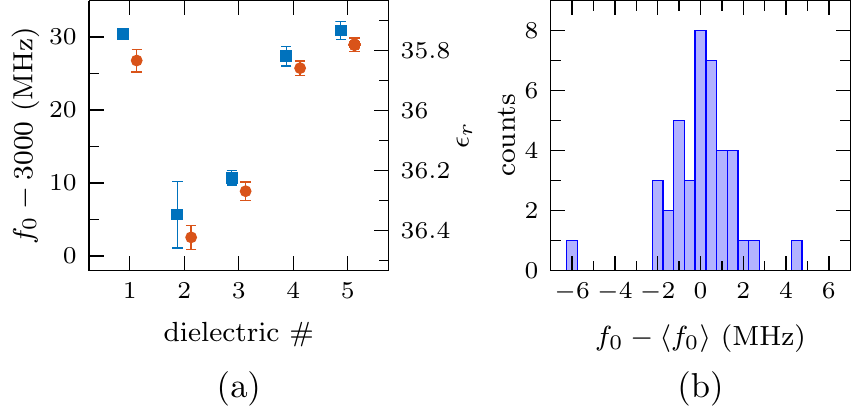}
\caption{(a) Resonant frequency for five different pieces of dielectric (in random order), each placed in one of two cavities. Each combination was assembled four times. The right $y$-axis shows the corresponding relative permittivity as estimated using simulations. (b)~Histogram of the deviations of all measured resonant frequencies from the average of each configuration.}\label{fig:reproducibility}
\end{figure}

The most important reason to use the screw cap design is to improve reproducibility in both fabrication and assembly. In order to test this, two cavities were fabricated, and five dielectrics accurately machined to a radius of $R_d=7.5$~mm were placed inside both cavities. Assembly of each combination was repeated four times. Figure~\ref{fig:reproducibility}(a) shows the resulting average resonant frequency and the standard deviation for each of the dielectrics, with the two cavities shown in different colors. The right axis shows the corresponding relative permittivity of the dielectrics estimated using simulations performed with CST Microwave Studio software~\cite{MWS} (MWS). The difference in the cavity housings is $2.4\pm0.9$ MHz, well within the tuning range. Unfortunately, different pieces of dielectric deviate more. It is currently unknown whether this difference arises from machining inaccuracies, damages or chips in the material, or differences in the relative permittivity of the dielectric. Figure~\ref{fig:reproducibility}(b) shows a histogram of the deviation from the average resonant frequency. This shows that assembling the cavity can be done reproducibly within a few MHz.

\begin{figure}
\includegraphics{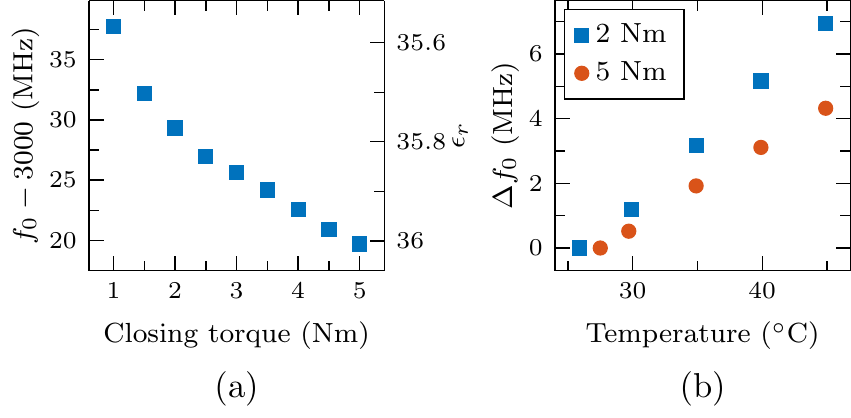}
\caption{(a) Resonant frequency as a function of the closing torque applied to the screw cap. (b)~Resonant frequency as a function of temperature for a cavity closed with two different torques.}\label{fig:stress}
\end{figure}

To test the influence of stress on the dielectric, the resonant frequency of one of these cavities was measured as a function of the closing torque applied to the cap. This is shown in Fig.~\ref{fig:stress}(a). This shows that with the application of stress on the dielectric, the frequency drops drastically. Control over the stress applied on the dielectric is therefore critical for a reproducible design.

Figure~\ref{fig:stress}(b) shows the change in resonant frequency as a function of the temperature of the cavity, performed at two different closing torques. For a copper vacuum cavity, the change is expected to be -50 kHz/K due to thermal expansion of the copper, whereas a change of $373\pm9$ kHz/K and $250\pm3$ kHz/K are found for a small and large stress respectively. Although the permittivity of the material changes with different temperature, this is typically only a few ppm, and is therefore not expected to have a large influence. The change is therefore attributed to the different thermal expansions of the dielectric and the housing, causing a change in stress on the material. Although this larger change in resonant frequency increases the tunability of the cavity through temperature, it also means that the requirements on thermal stability increase.

\begin{figure}
\includegraphics{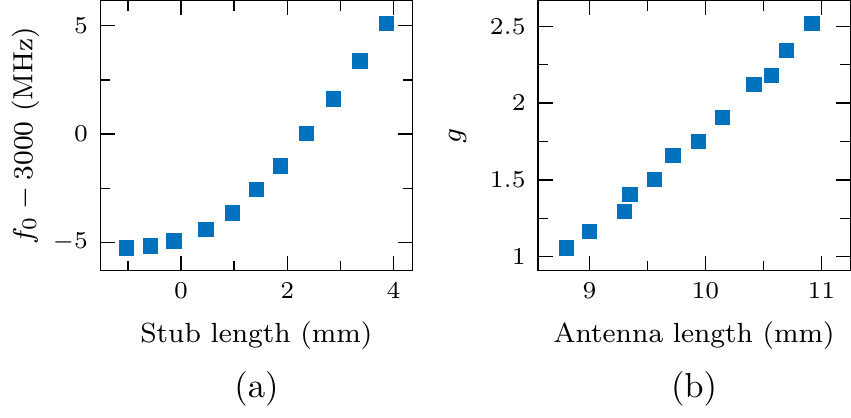}
\caption{(a) Resonant frequency as a function of stub depth. (b)~Coupling factor as a function of the antenna length.}\label{fig:tunability}
\end{figure}
Next, a cavity was designed to be implemented in a TEM. The two requirements on this cavity are that its resonant frequency must lie at 2.9985~GHz to allow for synchronization to a laser system, and that large powers can be coupled into the cavity. Shown in Fig.~\ref{fig:tunability}(a) and (b) are the resonant frequency and coupling factor of the cavity for varying stub and antenna lengths respectively. With the stub, the resonant frequency can be increased by up to $10.38\pm0.01$~MHz. With the antenna, the impedance of the cavity can be accurately matched to minimize power reflection.

\begin{figure}
\includegraphics{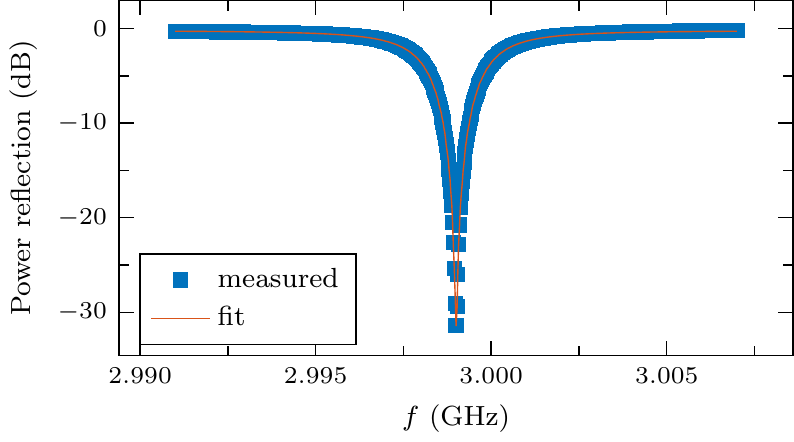}
\caption{Power reflection as a function of frequency, showing that good coupling can be reached at the desired frequency. From the fit, the quality factor of the cavity has been determined to be $(2.82\pm0.04)\times10^3$.}\label{fig:S11}
\end{figure}
Shown in Fig.~\ref{fig:S11} is the power reflection of the cavity tuned for insertion in a UEM, showing that the cavity can be tuned to accept power at the desired frequency. The absorbed power $P_\text{abs}$ as a function of frequency $f$ can be fitted with
\begin{equation}
\frac{P_\text{abs}}{P_\text{inc}}=1-|\Gamma_v|^2=\frac{4g}{(1+g)^2}\frac{1}{\frac{4Q^2}{(1+g)^2f_0^2}(f-f_0)^2+1}\,,
\end{equation}
with $P_\text{inc}$ the incoming power, and $f_0$ the resonant frequency. From the fit, the quality factor is determined to be $Q=(2.82\pm0.04)\times10^3$. Comparing this with simulations, the loss tangent of the material is then expected to be $\tan\delta=(2.38\pm0.05)\times10^{-4}$. However, the actual loss tangent is expected to be slightly lower, as the measured quality factor tends to be lower than simulations~\cite{ThesisThijs}, likely due to small imperfections.

The performance of this cavity inside a TEM has been reported in Ref.~\citenum{Verhoeven2018}. The same design was reproduced multiple times, allowing for the use of two TM$_{110}$ mode cavities synchronized to a TM$_{010}$ mode compression cavity, as reported in Ref.~\citenum{Verhoeven2018}.

\subsection{Dual mode cavity}
\begin{figure}
\includegraphics{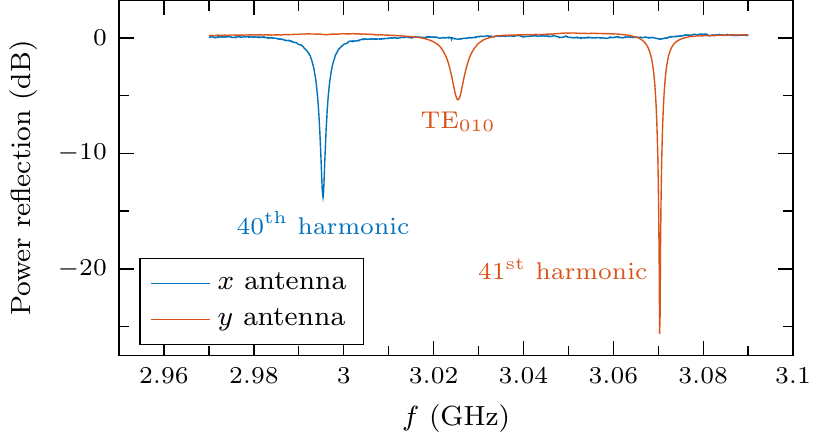}
\caption{Reflection parameter of a dual mode cavity as seen through the two antennas. Apart from the two perpendicular TM$_{110}$ modes a different mode is also visible, assumed to be a TE$_{010}$ mode.}\label{fig:dualModeSpectrum}
\end{figure}

\begin{figure}
\includegraphics[width=8.5cm]{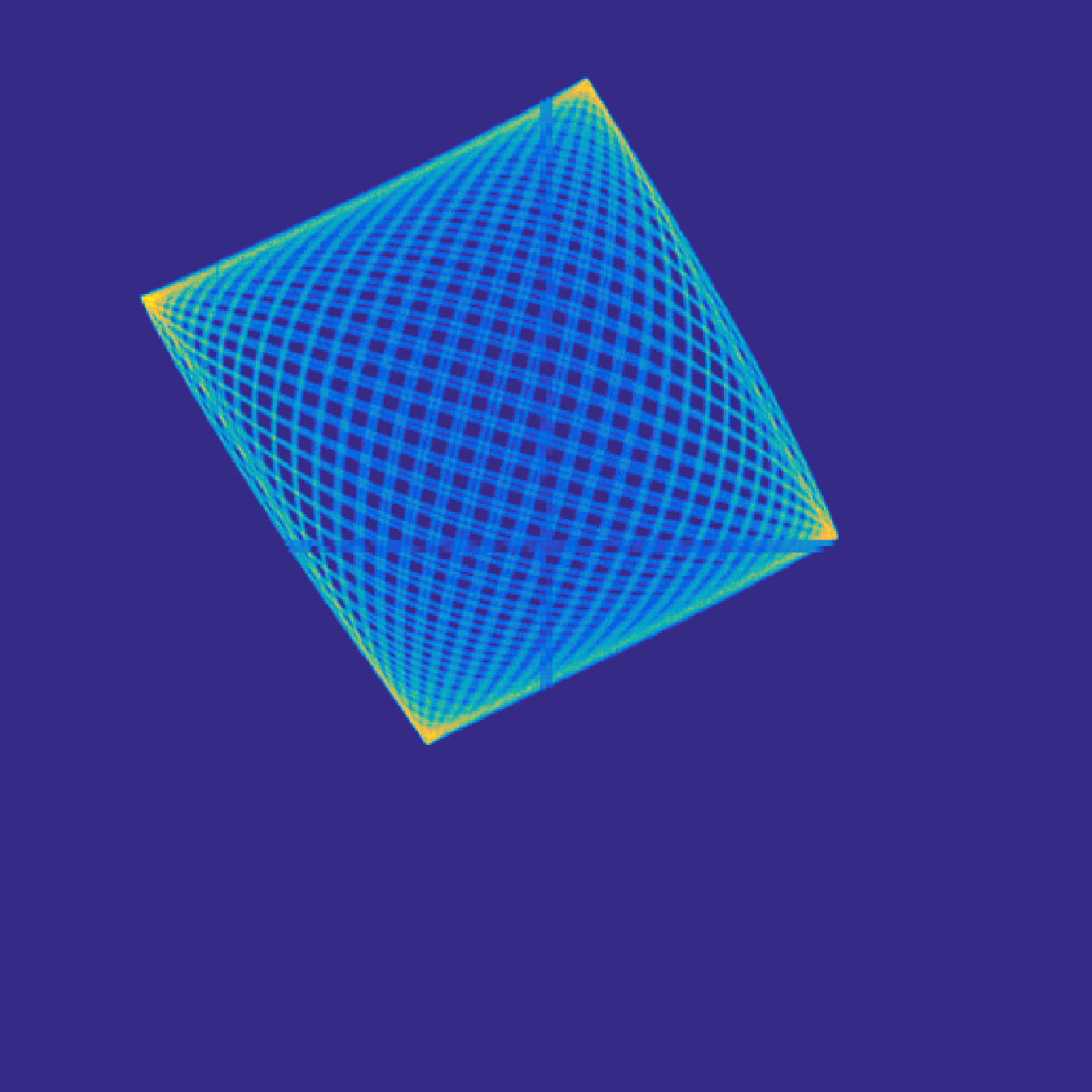}
\caption{Electron beam deflected onto a detector following the Lissajous pattern.}\label{fig:lissajous}
\end{figure}
Next, a dual mode cavity was designed. For a rotationally symmetric cavity, there is no preferential direction for the modes. Due to the presence of the antenna and stub this degeneracy is broken, fixing the mode orientation. Typically, a small transverse component remains, likely due to imperfections, giving rise to a slightly elliptical streak~\cite{ThesisAdam}. For the dual mode cavity, however, the cavity shape is modified to shift the frequency of one orientation. This is currently done by making the dielectric material elliptical and placing it in a circular housing, as shown in Fig.~\ref{fig:cavitySchematic}(b). Along the short axis of the dielectric, the frequency is then increased. This is done in such a way that the lower frequency can be synchronized to the 40$^\text{th}$ harmonic of a 75~MHz laser oscillator, and the higher frequency to the $41^\text{st}$ harmonic.

Figure~\ref{fig:dualModeSpectrum} shows the power reflection as a function of frequency for the two different antennas of a dual mode cavity. As can be seen, each antenna can strongly excite one of the two perpendicular modes, and couples a negligible amount of power into the other mode. Also present in this frequency range is a mode which is assumed to be a TE$_{010}$ mode. However, it is located several tens of MHz away from the TM$_{110}$ modes that its presence is not noticeable during operation.

Shown in Fig.~\ref{fig:lissajous} is the electron beam on the detector after passing through the dual mode cavity with power fed into both modes synchronized to the laser system. This shows that both can indeed be accurately mode-locked to result in a deflection following a Lissajous pattern. In Ref.~\citenum{VanRens2018} both the synchronization system and the performance in a TEM are presented of a dual mode cavity of the same design.

\subsection{Low loss cavity}
\begin{figure}
\includegraphics{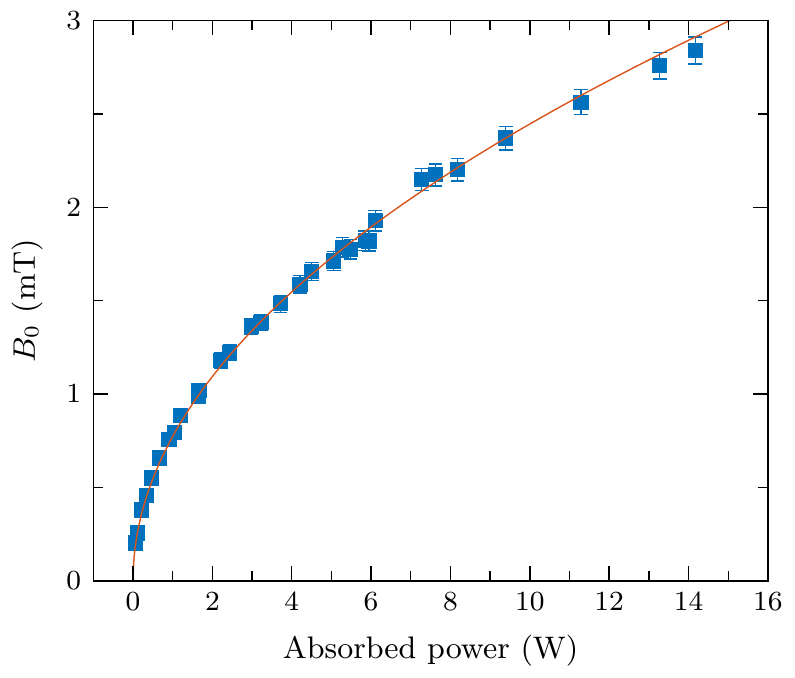}
\caption{Magnetic field strength as a function of input power of a cavity optimized for low power loss.}\label{fig:lowLossB}
\end{figure}
Finally, an optimized cavity design was manufactured and tested. For this, a dielectric with a radius of $R_d=6.25$~mm was made and placed in a cavity of radius $R_c=36.8$~mm. The magnetic field strength was then measured by deflecting a 30 keV electron beam onto a detector over a distance of $d=27$~mm. With the cavity length optimized for 30 keV electrons, the field strength can then be calculated from the length of the streak $L_\text{streak}$ on the detector using
\begin{equation}
B_0=\frac{\gamma m_e\omega}{4|q_e|d}L_\text{streak}\,.
\end{equation}
Shown in Fig.~\ref{fig:lowLossB} is the magnetic field strength of this cavity as a function of the input power. The solid line shows a fit of the data using $B_0\propto\sqrt{P}$, from which the power loss is found to be $P/B_0^2=(1.67\pm0.04)\times 10^6$~W\,T$^{-2}$. This is considerably lower than the value of $(18.0\pm0.1)\times 10^6$~W\,T$^{-2}$ that can be reached in an optimized vacuum cavity, showing the main advantage of using a dielectric material.

At a total input power of $P=14.2\pm0.2$ W a magnetic field of $B_0=2.84\pm0.07$~mT was achieved. Deflecting an electron beam over a 10~\textmu{}m slit with this cavity would result in pulses of $\tau=106\pm3$~fs at 30~keV, or $\tau=204\pm5$~fs at 200~keV. Table~\ref{table:lowLoss} summarizes the properties of the optimized cavity compared to the expectations from MWS simulations.

\begin{table}
\caption{Simulated and measured properties of the low-loss cavity. For the simulations, $\epsilon_r=35.55$ and $\tan\delta=1\times10^{-4}$ are used.}\label{table:lowLoss}
\centering
\begin{tabular}{|l|cc|}
\hline
&MWS&measured\\
\hline
$f_0$ (GHz) & 2.995 &$2.9998$\\
$Q$ & $6.00\times 10^3$ &$(5.02\pm0.04)\times 10^3$\\
$P/B_0^2$ ($10^6$ W\,T$^{-2}$) & 1.79 & $1.67\pm0.04$\\
\hline
\end{tabular}
\end{table}

\section{Conclusions}
A robust and reproducible design for a dielectric microwave cavity has been presented. Control over the resonant frequency of the cavity is crucial for synchronization to a laser system or other cavities. This design has therefore enabled all the experiments that will be presented in the rest of this thesis. Furthermore, the robustness of the cavities allows for better characterization of the dielectric material, enabling more advanced cavity designs. Two of these advanced designs were also presented in this chapter.

The first, called the dual mode cavity, supports two perpendicular modes at different well-defined frequencies. This allows for streaking the electron beam in a Lissajous pattern, extending the range of pulse repetition rates that can be achieved with a cavity. The second design was optimized for low losses. With this design, losses were reduced to $P/B_0^2=(1.67\pm0.04)\times10^6$~W\,T$^{-2}$, and field strengths up to $B_0=2.84\pm0.07$~mT were demonstrated.

The reproducibility is limited by the dielectric material, which showed by far the largest deviations for currently unknown reasons. The tuning range achievable with the combination of a stub, the controlled application of mechanical stress and the temperature is large enough to correct for these deviations. Alternatively, each dielectric can be characterized in a test cavity, after which a housing can be custom made.

\section*{Acknowledgement}
This work is part of an Industrial Partnership Programme of the Foundation for Fundamental Research on Matter (FOM), which is part of the Netherlands Organisation for Scientific Research (NWO).

\bibliography{bibFile}
\end{document}